# Light-effect transistor (LET) with multiple independent gating controls for optical logic gates and optical amplification

Jason K. Marmon[1-3], Satish C. Rai[4], Kai Wang[4], Weilie Zhou[4], and Yong Zhang[1-3]

[1]Nanoscale Science Program, University of North Carolina at Charlotte, Charlotte, NC 28223, USA.

[2]Department of Electrical and Computer Engineering, University of North Carolina at Charlotte, Charlotte, NC 28223, USA.

[3]Center for Optoelectronics and Optical Communications, University of North Carolina at Charlotte, Charlotte, NC 28223, USA.

[4]Advanced Materials Research Institute, University of New Orleans, New Orleans, LA 70148, USA.

Correspondence and requests for materials should be addressed to Y.Z. (e-mail: yong.zhang@uncc.edu) and W.Z. (e-mail: wzhou@uno.edu).

**Abstract:** Modern electronics are developing electronic-optical integrated circuits, while their electronic backbone, e.g. field-effect transistors (FETs), remains the same. However, further FET down scaling is facing physical and technical challenges. A light-effect transistor (LET) offers electronic-optical hybridization at the component level, which can continue Moore's law to quantum region without requiring a FET's fabrication complexity, e.g. physical gate and doping, by employing optical gating and photoconductivity. Multiple independent gates are therefore readily realized to achieve unique functionalities without increasing chip space. Here we report LET device characteristics and novel digital and analog applications, such as optical logic gates and optical amplification. Prototype CdSe-nanowire-based LETs show output and transfer characteristics resembling advanced FETs, e.g. on/off ratios up to ~$1.0 \times 10^6$ with a source-drain voltage of ~1.43 V, gate-power of ~260 nW, and subthreshold swing of ~0.3 nW/decade (excluding losses). Our work offers new electronic-optical integration strategies and electronic and optical computing approaches.

**Text:** As basic electronics building block, a field-effect transistor's (FET's) primary switching function is widely used in both logic and memory chips. A typical FET is a three-terminal device consisting of source (*S*), drain (*D*), and gate (*G*) contacts – where the *S-D* conductivity is modulated to realize on and off states by applying a voltage or an applied electric field through *G*.[1] Although FETs have evolved structurally from early planar to current 3D geometries in parallel with the continual shrinkage of its lateral size, the basic operation principle remains the same, which leads to ever greater fabrication complexity, and ultimately to challenges in gate fabrication and doping control.[2-6] Various new technologies, such as FinFET,[2, 7] and tunnel-FET,[8] have been developed in recent years to enable the continuation of Moore's law,[9] but further development with current technologies are uncertain.[10] Other options are being explored as alternatives, which include semiconductor nanowire (SNW) based FETs,[11-13] FETs comprised of 2D materials,[14, 15] and FETs with sophisticated gate structures,[16] such as multiple independent gates[3, 5] or a gate with embedded ferroelectric material.[17] There is, however, no clear pathway for overcoming a FET's intrinsic physical limitations[18-20] dictated by its operation mechanism, such as random dopant fluctuations[6] and gate fabrication complexities,[21] and no viable rival technology currently exists. We offer a competitive alternative with additional unique functionalities. The light-effect transistor (LET) is a two-terminal device composed of a metal-semiconductor-metal (M-S-M) structure, where each M-S junction serves as either the *S* or *D* contact, and the two contacts are separated by a semiconductor nanostructure channel. Fig. 1 contrasts SNW-based LET and FET structures to reveal the apparent structural simplicity offered by a LET – no physical gate is required. A LET's operation mechanism is distinctly different from a FET in two regards: (1) the *S-D* conductivity is solely modulated by light or an optical frequency electromagnetic field, which contrasts a FET's electrostatic control through an applied DC voltage, and (2) current carriers are generated through optical absorption rather than by thermal activation of dopants. In other words, a LET employs optical gating based upon the well-known photoconductive mechanism[22] that has typically been of interest in photo-detection. Inherent advantages stem from a LET's simplistic architecture, which include (1) eliminating gate fabrication complexity, and (2) avoiding difficulties with doping control. These attributes remove the two primary challenges or intrinsic limitations facing conventional FETs in regards to their continuous down scaling to the quantum regime,[23] and in the meantime, they offer the potential for reduced fabrication costs. While a LET's most basic application emulates a FET when it operates under one-beam illumination as in a photo-detector, it offers functions not readily achievable by either a FET[24] or a photo-detector,[25, 26] when it operates differently than a typical photo-detector, e.g., when responding to multiple independent light beams.

Light-induced electrical conductivity change is a well-known phenomenon typically used for photo-detection. In fact, SNW devices structurally similar to our LET have been investigated as photo-detectors.[25, 26] At first glance, it may appear that a LET simply employs a photo-detector's switching function to emulate a FET. In reality, most photo-detectors lack desirable FET-like characteristics and are therefore unsuitable for LET use. It is therefore important to understand the differences between a photo-detector, LET, and FET to appreciate the LET's novelty. Photo-detection typically relies upon a p-n junction-based device, because it usually offers superior performance over a simpler M-S-M device based on the photoconductive mechanism. This arises from the M-S-M structure typically requiring a larger bias to drive carriers through the S region.[1, 22] Significantly, a p-n junction based photo-detector has a distinctly different I-V characteristic under illumination from a photoconductive based one, and only the latter can offer a light I-V resembling that of a FET with gate voltage on. The photoconductive mode's disadvantage is eased through reduced device dimensions, as demonstrated by SNW-based photo-detectors,[25, 26] and the LET application in this work. Its structural simplicity should provide further advantages at the genuine nanoscale. We note that photo-detector structures that are difficult to dope may also employ a M-S-M structure.[1, 22] Therefore, a LET does not employ a new device structure or mechanism. Its novelty stems from its stringent electrical and optical characteristics that can (1) replicate the basic switching function of the modern FET with competitive (and potentially improved) characteristics, and (2) enable new functionalities not available in modern FETs, as well as, allow new applications beyond those offered by conventional photo-detectors. While under single-beam illumination, a LET requires a high on/off ratio under optical gating, which resembles a FET under gate-voltage control or a photo-detector with high photoconductive gain. Despite this similarity, a LET should also be characterized for a pertinent FET parameter known as "subthreshold swing", which measures how much gate action is required to turn on the device, and is normally not of interest in photo-detection applications. Under simultaneous multi-beam illumination, which is usually irrelevant for photo-detection, the multiple independent gating capability enables a LET to demonstrate previously unreported functions, such as optical logic (*AND* and *OR*) gates and optical amplification as an analog application. In contrast, multiple independent gating has been a very challenging task for FETs.[3] These unique functionalities are of great interest for optical computing and novel optical detectors. To summarize, LET novelty, in comparison to photo-detectors, is two-fold: First, LETs are characterized electrically in a very different manner than photo-detectors, as photo-detectors are not typically explored for electronic functions found in a FET. Second, LETs utilize their multi-beam response while a photo-detector does not. In comparison to FETs, a LET's gating mechanism is distinctly different from a FET's, which makes easy the LET's multi-gate capability, and allows a LET to offer functions beyond those in a typical FET. Furthermore, a LET's frequency response or switching speed is limited by the carrier lifetime in its conducting channel. While this effect is shared with a FET, a FET's response is limited by its gate capacitance.

Our LET devices employ readily available CdSe SNWs.[27, 28] Fig. 2 provides material and device characteristics. Fig. 2A displays an SEM image of a 10-µm-long CdSe SNW (device 1 or D1) with indium (In) contacts forming M-S junctions at each end. The uniform single-crystalline CdSe SNW was grown in wurtzite phase along the [0001] axis with a diameter of ~80 nm, as revealed by the low magnification transmission electron microscopy (TEM) image in Fig. 2B, with the selected area diffraction pattern (SADP) as inset, and Fig. 2C's high-resolution TEM (HRTEM) image showing a 0.69 nm inter-planar spacing. The gold catalyst at the SNW end (Fig. 2B) suggests the vapor-liquid-solid growth mechanism.[29] The CdSe-SNW's laser-power-dependent photoluminescence (PL), Fig. 2D, shows a strong emission peak at 1.78 eV that matches CdSe's bandgap energy.[30] The inset overlays a PL map upon an optical image to demonstrate relatively homogenous SNW emission, and by extension, homogenous material quality across the SNW channel. In Fig. 2E, the output characteristic, S-D current $I_{ds}$ vs. *S-D* voltage $V_{ds}$, is demonstrated for the device with and without light illumination using a halogen light, where illumination optically modulates or "gates" the electrical conductivity between dark ("off") and illuminated ("on") states. The $I_{ds}$ *vs.* $V_{ds}$ curves of the two states clearly resembles those of a FET's off and on states,[1] respectively, especially when $V_{ds} < \sim 7$ V.

The novel LET requires performance metrics for evaluation and comparison against FETs; thus, FET figures of merit are adapted, such as the two important input-output relationships: (i) "*output characteristics*" or $I_{ds}$ *vs.* $V_{ds}$ under a constant illumination condition $P_g(\lambda_g)$, which is equivalent to the FET's output characteristic under a constant gate voltage $V_g$; and (ii) "*transfer characteristics*" or $I_{ds}$ *vs.* $P_g(\lambda_g)$ under a constant $V_{ds}$, which is equivalent to a FET's $I_{ds}$ *vs.* $V_g$ under a constant $V_{ds}$. A FET's gate voltage, $V_g$, is replaced by a LET's gate power $P_g(\lambda_g)$, which not only serves the same function of modulating *S-D* conductivity as in a FET functions but also offers an avenue to achieve additional novel functions. Characteristic (i) is shared by both LET and photo-detection applications, while characteristic (ii) is required for LETs and FETs as a measure of turn-on energy, and in particular for LET to realize the novel functions.

LETs – being electrical-optical hybrid devices where an electrical field, $V_{ds}$, and an optical field, $P_g$, together modulate the electrical output, $I_{ds}$ – differ from other optoelectronic devices, such as light-emitting devices (or solar cells) that require an electrical (or optical) input to generate an optical (or electrical) output. LET features far greater gating flexibility and ability than a FET. Optical gating through $P_g(\lambda_g)$ has two basic control parameters: wavelength, $\lambda_g$, and power level, $P_g$, under one-beam CW operation, but it can be readily extended to other operation modes, for instance, multiple independent beams and pulsed illumination represented as $P_g(\lambda_{g1}, \lambda_{g2},...,\lambda_{gN})$ and $P_g(t,\lambda_{g1}, \lambda_{g2},...,\lambda_{gN})$, respectively. In this work, we first fully characterize LET output and transfer characteristics under one-beam CW operation with two illumination conditions: (a) illuminating the center of the SNW with a focused CW laser ("focused illumination") with an optical diffraction-limited spot size at wavelengths of 633, 532, 442, or 325 nm; and (b) illuminating the LET uniformly with "white light" from a halogen lamp ("uniform illumination"). Results for two devices, device 1 (D1) and device 2 (D2) with lengths of ~10 and ~5.5 µm and similar diameters (~80 nm), are presented to illustrate general LET properties, and to demonstrate the potential for characteristic tuning and optimization. The two devices were fabricated in essentially the same way. We then explore some unique functionalities under two-beam illumination that are not readily achievable with a typical FET.

Device dark currents reveal negligible reverse bias current and rectification (diode-like behavior) under forward bias, e.g. Fig. 2E. LET operation occurs under forward bias for both devices. Rectification is indicative of asymmetric In/CdSe contacts for both devices, where one M-S junction is close to ohmic and the other forms a Schottky contact;[31] large asymmetric contacts are desired as they drastically reduce the dark current or off state and thereby improve the on/off ratio. The Schottky barrier largely determines the turn-on voltage, $V_{D,on}$, which is ~8 V for D1 and > 21 V for D2. For instance, D2 shows nearly resistive behavior up to $V_{ds} = 21$ V with $I_{ds}$ reaching only ~15 pA, compared to D1's range from ~1 nA to ~4 µA over 1-21 V. The vast difference between the two devices might stem from a thin $SeO_x$ layer[32] (x = 2-3) at the In/CdSe junction, although the details require further study. These results hint that dark or off state parameters can be controlled through M-S contact engineering.

Representative LET output characteristics are shown in Figs. 3A-D for D1 and in Figs. 3E-F for D2, respectively, to exemplify how LET performance depends on the gate power/wavelength, illumination condition, and device variation. D1 exhibits two well-separated plateaus, respectively, starting at $V_{ds}$ ~ 4-5 V and ~14-18 V depending on the gate wavelength and power. For example, the second plateau's onset is at ~14-15 V for 633 nm illumination but shifts to ~16-18 under 442 nm excitation. Two tunable plateaus can potentially offer two distinct, customizable on states. For D2, the first and second plateau are comparatively not well separated, and both 532 nm and halogen illumination have their first plateau at ~2 V with respective power-dependent, second plateaus at ~6-7.5 V (532 nm) and ~5-5.75 V (halogen). Each plateau appears at respectively lower $V_{ds}$ values than in D1, and because of the extremely low dark current, the long second plateau extends to the highest $V_{ds}$ measured. For D1, the maximum on/off ratios typically occur at $V_{ds} < 5$ V, and vary from $10^2$ to $10^4$ depending on the gate power and wavelength. For instance, Fig. 3B contains on/off ratios of ~5x$10^4$ and ~2x$10^4$ at $V_{ds}$ = 1.43 and 4.95 V, respectively, when $P_g(532nm) \approx 2$ µW. The on/off ratios for D2 in Fig. 3E are ~1.0x$10^6$ and ~1.1x$10^6$ at $V_{ds}$ = 1.43 and 4.95 V when $P_g(532nm) \approx 2.6$ µW. When $P_g(halogen) \approx$ 69 µW in Fig. 3F, the on/off ratios are ~6x$10^5$ and ~1x$10^6$ at $V_{ds}$ = 1.43 and 4.95 V, respectively. Differences between D1 and D2 indicate that a LET's characteristics may be tuned and optimized through material and device engineering. A large M-SNW contact barrier is generally desirable for producing small off state currents over the operation range, and can be optimized to maximize the on/off ratio. Note that current levels for different "gate" wavelengths in Figs. 3A-D showed considerable variations, which is fundamentally due to wavelength-dependent light-matter interaction effects, e.g. absorption and carrier dynamics, and illumination conditions, e.g. power density and beam size. This feature offers the unique LET advantage of flexibility in achieving gate functions compared to FETs.

The $I_{ds}$-$V_{ds}$ curves in Fig. 3 may be understood qualitatively with the photo-conductivity model proposed by Mott and Gurney.[22] The first plateau is associated with the "primary photoconductivity" which produces current as a result of photo-generated electrons and holes flowing through the nanowire under applied bias. A steady state condition is formed when just enough external carriers entering the nanowire through the electrodes replenish those leaving the device. The collection efficiency, $\Psi$, can be approximately described by $\Psi = w/L\ [1-\exp(x_0/w)]$, where $w$ is the carrier's mean free path (which is proportional to the applied field), $L$ is the nanowire's length, and $x_0$ is the illumination site measured from the anode. This theory suggests a continuous photocurrent increase from $V_{ds} = 0$ until saturation occurs at a sufficient $V_{ds}$ to produce $w \gg L$ (Figure S1 for simulated $\Psi$ vs. $V_{ds}$ curves). If all photons are absorbed, $\Psi$ is equivalent to the quantum efficiency, $\eta_{QE}$, defined as $I_{ph}/(eN_{ph})$, where $I_{ph}$ is the photo-induced current, and $N_{ph}$ is the number of absorbed photons. When current saturation occurs, $\eta_{QE} = 100\%$. For instance, absorbing 2 µW of 620 nm light with $\eta_{QE} = 100\%$ yields a 1 µA current. As $V_{ds}$ approaches $V_{D,on}$, a major Schottky barrier reduction[31, 33] allows excess carriers to enter the nanowire through the electrode, which then produce a drastic $I_{ds}$ increase that allows $\eta_{QE} \gg 1$. This simple picture is most applicable to D1. For D2's much higher M-SNW barrier, the two plateaus could respectively reflect photoconductive electron and hole contributions.

The transfer characteristics allow extraction of a few performance metrics. A FET's threshold gate voltage, $V_T$, and subthreshold swing, $S$, are respectively defined as the onset of a linear region in the $I_{ds}$-$V_g$ curve (i.e. voltage-controlled resistor behavior), and as the inverse linear slope on a semi-log $I_{ds}$-$V_g$ plot.[2] Their physical interpretations, respectively, are the gate voltage required for device operation and the gate voltage increment to induce an order of magnitude current change below $V_T$. A small $S$ value implies a small energy or power consumption to turn on or operate a FET. Fig. 4 contains D1's and D2's transfer characteristics, which in general, resemble a FET's transfer characteristics, e.g. increasing $I_{ds}$ as the gate power $P_g$ increases under constant $V_{ds}$, except a LET replaces $V_g$ with $P_g$. A LET's threshold gate power, $P_T$, then, corresponds to the onset of a linear $I_{ds}$-$P_g$ region for a given $\lambda_g$, and $S_{LET}$ is its subthreshold swing. Significantly, FETs usually do not operate in the "subthreshold swing" region, while a LET can employ this range to realize optical logic gates and for an interesting optical amplification effect. Taking D2's $I_{ds}$ vs. $P_g$ curves, Fig. 4E, with $\lambda_g$ = 532 nm as examples, typical $P_T$ and $S_{LET}$ values at $V_{ds}$ = 1.43 (4.98) V are, respectively, ~30. (~30.) nW, and ~2.8 (~2.5) nW/decade. For reference, advanced FETs have respective $V_T$ and $S$ parameters of 100-200 mV, and ~70-90 mV/decade.[34] At $V_{ds}$ =1.43 V, $P_g$ = 0.11 µW yields $I_{ds} \approx$ 0.35 µA, and a LET dynamic power consumption of ~0.5 µW, which is comparable to advanced FETs.[35] A LET's off-state energy consumption can be very low. For instance, the dark current is ~1 pA at $V_{ds}$ = 1.43 V with a corresponding static power consumption of ~1.5 pW, which is lower than a FET of similar length.[35] Significantly, the collection efficiency is expected to improve drastically at low $V_{ds}$ with nanometer-length devices (Figure S1), which should further reduce the static power consumption and provide lower $V_{ds}$ than those demonstrated here. The maximum applied laser power is about 3 µW and corresponds to a power density of ~0.60 W/mm$^2$, which is less than that delivered by an efficient light-emitting diode.[36] The actually used gate power is only about 10% of the applied power because the laser spot size is considerably larger than the nanowire diameter (Supporting Information for energy loss estimates). Reducing the beam size closer to the SNW's diameter could reduce $P_g$ by at least a factor of 10,[37] and, as is well established in FET devices, reducing the channel length can reduce required the $V_{ds}$ (Figure S1). Enhanced efficiency and reduced energy consumption could significantly reduce thermal issues plaguing nanoscale FET-containing electronics devices. We note that FET has a thermal dynamic limit of S ≥ (kT/q) ln(10) = 60 mV/decade at 300 K, whereas for LET, $S_{LET}$ is extrinsic in nature through the dependence of $w$ on the carrier density, which in turn depends on the defect density. Thus, $S_{LET}$ can be significantly improved by shortening conduction channel and perfecting the material quality.

The LET transfer characteristics are used to illustrate the underlying principles for a few important applications. D2's 532 nm illumination characteristics, Fig. 4E, are re-plotted on a double log scale in Fig. 5A, with only $V_{ds}$ = 1.43 and 4.98 V shown for clarity, to more clearly portray the three major operating regions: super-linear (dark gray region), linear (medium gray), and sublinear/saturation (light gray). Different regions can offer different unique applications, as the examples highlighted below.

(1) *AND logic gate and Voltage amplifier:* Fig. 5B demonstrates single beam illumination as a hybrid *AND* logic gate, which replicates the most basic FET logic function,[1, 38] using electrical input A = $V_{ds}$ and optical input B = $P_g$ with output denoted as *AxB*. This is achieved when $V_{ds}$ = 5V and $P_g$ is modulated between 0 and 2.60 µW.

One-beam operation could also act as a current source or voltage amplifier when operating in the output characteristic's saturation region, or even when utilizing a LET's two distinct on states (e.g. the first and second plateaus in Fig. 3B) to realize two-level logic gate and voltage amplifier functions.

(2) *Multi-independent-gate capability:* An important LET advantage is multi-independent-gate operation, where optical gates do not increase device dimensions. As an example, two-beam operation is demonstrated with independently controlled uniform illumination of halogen light and focused illumination of 532 nm laser, denoted as $P_{g1}$ and $P_{g2}$ respectively. Illumination by either single light beam produces its corresponding transfer characteristics, e.g. $I_{ds}$ vs. $P_g$ in Fig. 4, while two-beam illumination results in a 3D $I_{ds}$ vs. ($P_{g1}$,$P_{g2}$) plot (Figure S2A). However, the two-beam response fundamentally reflects the linearity of the single-gate response shown in Fig. 5A. To more clearly show this effect, a current enhancement factor $R$ is introduced by converting $I_{ds}(P_{g1},P_{g2})$ to $R(P_{g1},P_{g2})$, where $R = I_{ds}(P_{g1},P_{g2}) /[ I_{ds}(P_{g1}) + I_{ds}(P_{g2})]$. Figure S2A's data were converted with this definition and the corresponding R values are displayed in Fig. 5C's contour plot. Using the LET response characteristics in Fig. 5A and Fig. 5C, we demonstrate a few distinctly different LET functions that are not readily achievable using a FET, and can be realized with a single LET device. Figs. 5D-G demonstrate dual gate applications in three important $R(P_{g1},P_{g2})$ regions illustrated in Fig. 5C.

(2a) *Optical amplification:* This occurs in Fig. 5A's super-linear or subthreshold swing region and yields a region of $R >> 1$ in Fig. 5C, for instance, $R \approx$ 9-11. Fig. 5D yields single beam induced currents of $I_{ds,\ 532nm} \approx$ 11 nA (dark cyan line) and $I_{ds,\ halogen} \approx$ 37 nA (orange line), while simultaneous illumination produced ~11 times their sum with a $I_{ds,\ 2beam} \approx$ 525 nA (royal blue line). If the laser beam is viewed as a weak optical signal to be measured, and the halogen light (~1.6 µW) as a gate signal, an amplification factor of $m \approx 48$ is obtained. Optically-induced amplification of a LET's electronic signal replicates three-terminal phototransistor function, e.g. a bipolar transistor with a semi-transparent electrode,[39] where a small base-emitter bias leads to photo-current amplification. This feature may find broad application in weak optical signal detection.

(2b) *Optical AND logic gate*: Results shown in Fig. 5D can also be used for important optical logic operations, such as that in Fig. 5E. Two individually applied optical gates, with inputs of *A* and *B* respectively, produce two low current or off states represented as *(1,0)* or *(0,1)* in addition to the *(0,0)* off state (not shown for clarity). Only under simultaneous illumination does output *C* produce the on or *(1,1)* state. LET-enabled optical logic operations could lead to new optical or quantum computing approaches.[40]

(2c) *Optical summation*: Sum operations can be realized in Fig. 5C's linear response region, e.g. $R = 1$, as illustrated in Fig. 5F. In this figure, $P_{g1, 532nm}$ and $P_{g1, halogen}$ generate two independent signals of 2.00 and 0.32 µA, while simultaneous illumination produces a current of 2.43 µA or approximately their numerical sum. This region is convenient for producing multiple states, such as for memory devices.

(2d) *Optical OR logic gate:* Current saturation is achieved when $R = ½$, and can function as an optical *OR* logic gate, Fig. 5G. When $A = P_{g1}(532nm) = 0.63$ µW and $B = P_{g2}(halogen) = 69.1$ µW, individual illumination as *(1,0)* and *(0,1)* states or dual illumination as the *(1,1)* state all produce comparable $I_{ds}$ values; all three on states contrast the off state of pA-level $I_{ds}$ denoted as *(0,0)* (not shown for clarity). A single LET could perform more complex logic functions concurrently by combining $V_{ds}$ control with dual optical gate ability, such as a three-terminal

*AND* gate with output *AxBxC*, or with simultaneous *AND* and *OR* gates with *Ax(B+C)* output. Truth tables for these logic operations and their proposed symbols are provided (Figure S4). Significantly, LET's do not require multiple switches or single-NW devices to realize complex logic functions, which could require fewer devices to perform identical or enhanced functionality. Thus LETs offer an additional pathway for achieving high device densities on a single chip.

(3) *Differentiator and Optically gated phase tuner*: Complementary to the above mentioned functions, LETs can also be used as a differentiator under zero or low $P_g$, and as a phase tuner as $P_g$ is increased. Fig. 5H shows the $I_{ds}(t)$ vs. $V_{ds}(t)$ curves for different $P_g$ values, where $V_{ds}(t)$ is a sine wave modulation with an amplitude of 5.0 V and a DC offset to remove the negative portion. The $I_{ds}(t)$ curve exhibits a 90° phase delay with respect to $V_{ds}(t)$ when $P_g = 0$, which indicates that the device functions as a differentiator by converting a sine wave into a cosine wave; increasing $P_g$ results in a tunable phase shift that gradually approaches zero, e.g. at $P_g = 2.6$ μW. This effect can be understood as changing the LET's impedance by varying the gate power.

Finally, we remark on LET quantum scale operation. A LET's structural simplicity removes potential obstacles that FETs face for further down scaling. A LET shares the same limit of a FET, that is, the nanostructure dimensions practically achievable, e.g. 1-7 nm for Si nanowires,[41] but LETs do not require complex and sophisticated fabrication steps for physical gates and doping. In general, ballistic transport theory suggests that commercially viable currents could be achieved in quantum structures.[42] Quantum conductance, which limits 1-D ballistic transport, is given by $G = nG_0$, where $G_0 = 2e^2/h$ is the minimum conductance and *n* are integers representing quantized energy levels. This equation yields a maximum quantum impedance for the conducting channel of $Z_0 = 1/G_0 = 12.9$ kΩ. Given the highly localized nature of the 1-D energy density of states, LET conductivity is expected to be quantized, and thus, tunable using different photon energies. Industry may employ at least two basic illumination modes in an integrated LET circuit, depending on the application: (1) uniform, broad-area illumination over a high-density LET array with SNWs, or (2) separated light beams directed to individual or small groups of LETs through, for instance, sharp fiber tips or a small emitter embedded on the same chip. For either mode, multiple light sources of the same or different wavelength(s) and/or intensities can be combined into one beam but controlled independently.

In summary, the LET concept represents a drastically different approach for FET-based IC technologies by using an all optical, rather than a physical gate mechanism. A LET explores the well-known photoconductivity attribute of a semiconductor that is naturally and commonly used for photo-detection. Here, we demonstrate digital and analog applications typically only achievable with transistors, as well as, functions that FETs cannot achieve. Most significantly, the LET gate function can provide much greater flexibility than a FET, including tunable gate properties and multiple independent gates. Notably, a LET can continue Moore's law without the FET complications and limitations associated with gate fabrication and doping control through: (1) a simple device architecture to potentially reduce fabrication costs; (2) feasible down scaling to the quantum level; (3) efficient, multi-functional ability in a single device; and (4) operation at low optical gate power(s), which negate thermal issues plaguing nanoscale electronics devices. The general LET operation principle is independent of a particular material system, thus, when applied to silicon, the existing silicon-based microelectronic and photonic technologies can be readily adopted by LET technology. The LET concept can also be extended to develop other light-effect devices.

**Acknowledgments:** This work was partially supported with funds from Y.Z.'s Bissell Distinguished Professorship at UNC-Charlotte. We extend deep appreciation and thanks to Dr. Quiyi Ye for valuable discussions and comments. J.K.M. thanks John "Jack" Krause for useful electronics-related discussions. S. C. R. thanks Shuke Yan for assistance in device fabrication.

**Author Contributions:** Y.Z. and W.L.Z. guided work at UNCC and UNO respectively. J.K.M. conducted the optical and electrical measurements and analyzed the data, S.C.R. grew the CdSe nanowires and fabricated the devices, and K.W. collected the TEM data and created the 3D schematics in Fig. 1. Y.Z. and J.K.M. wrote the manuscript. All authors have reviewed, edited, and commented on this manuscript.

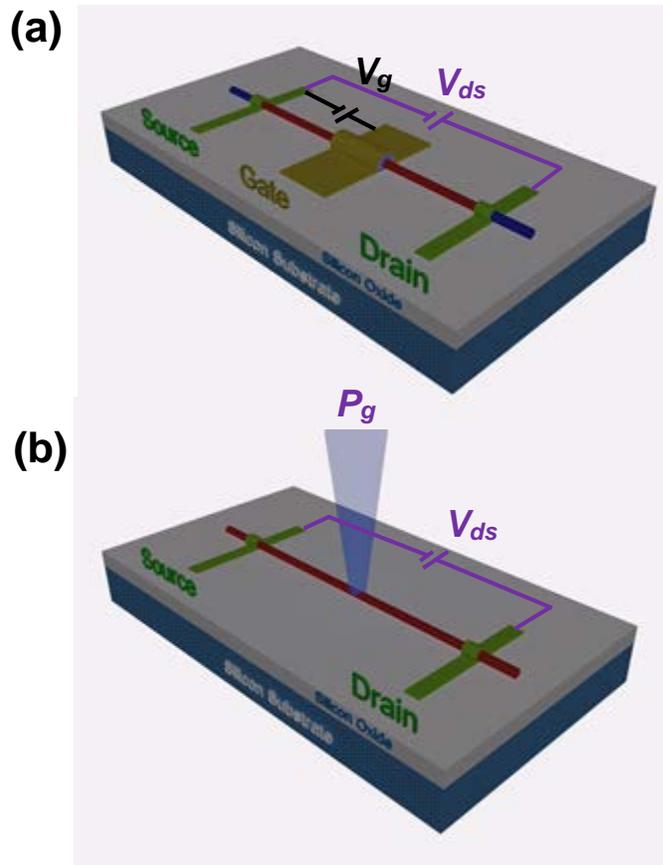

**Figure 1** Schematic comparison between a semiconductor-nanowire-based (SNW-based) field-effect transistor (FET) and a light-effect transistor (LET). **(a)** A FET is a three terminal device where the source-drain, *S-D*, current, is driven by an *S-D* voltage and may be modulated through a gate (*G*) voltage applied through its *G* contact. **(b)** A LET is a two terminal device where the *S-D* current is modulated with one or multiple independently controlled light beams fused together through an optical combiner. Color codes are SNWs in red, *S* and *D contact in green*, *G* contact in yellow, and the gate dielectric (under the *G* contact) in blueish-gray. The blue-colored SNW tips past the *S* and *D* contacts indicate different FET doping types along the conducting SNW channel. It is assumed that these devices are resting upon an insulating substrate.

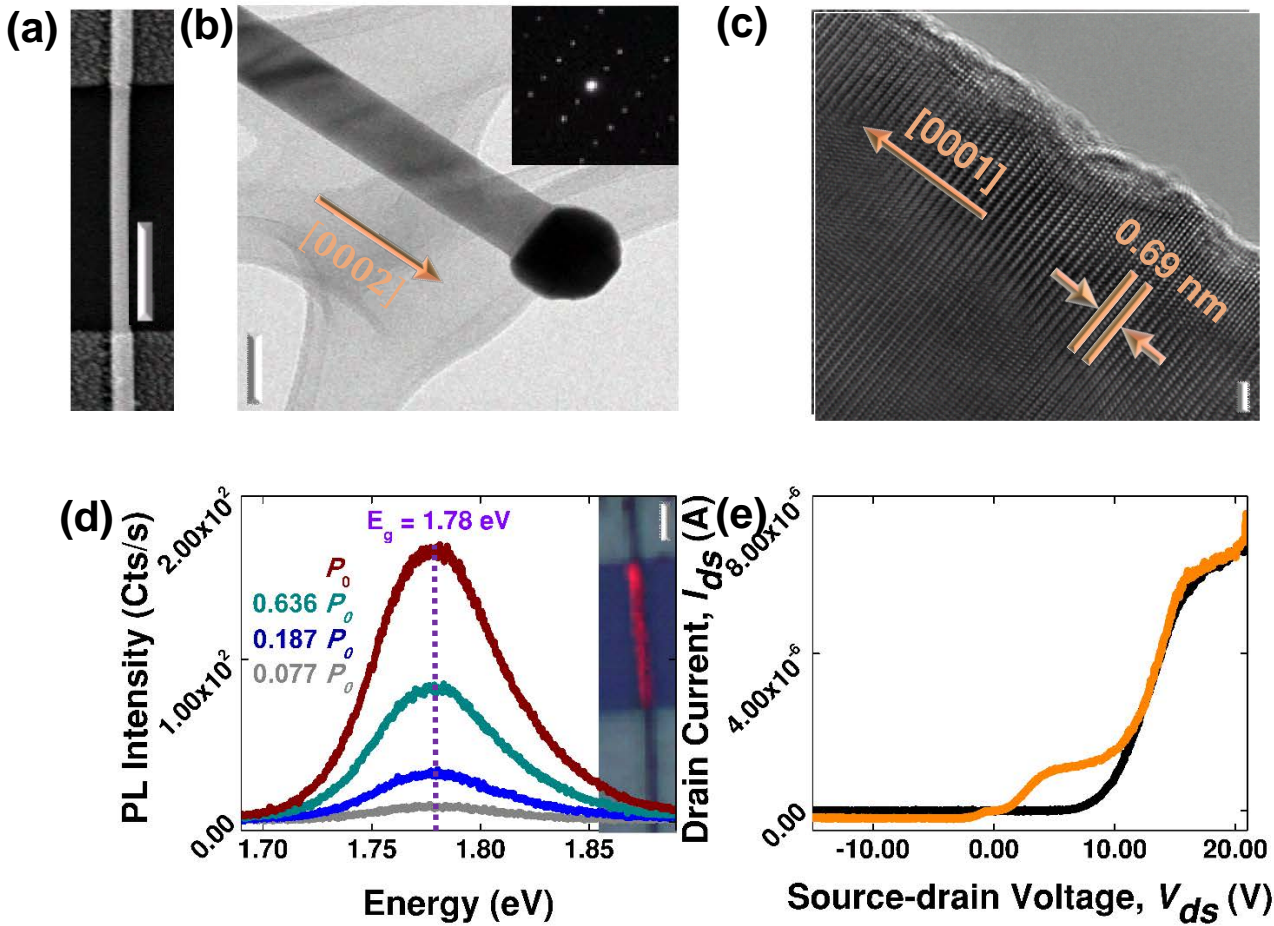

**Figure 2** LET characterization. **(a)** SEM image of a typical In-CdSe-In device (2 µm scale bar). **(b)** TEM (100 nm scale bar) with SAED inset, and **(c)** HRTEM image (2 nm scale bar) of a representative CdSe nanowire. The TEM results indicate single crystalline CdSe with well-ordered lattice plane spacing of 0.69 nm along the [0001] growth direction. **(d)** PL spectra obtained under 442 nm excitation at different powers ($P_0$ = 1.5 µW). Inset contains PL map overlaid upon an optical image of D1 (4 µm scale bar). (E) Source-drain current, $I_{ds}$, as a function of source-drain voltage, $V_{ds}$, under dark (black line) and halogen light illumination (orange line) conditions.

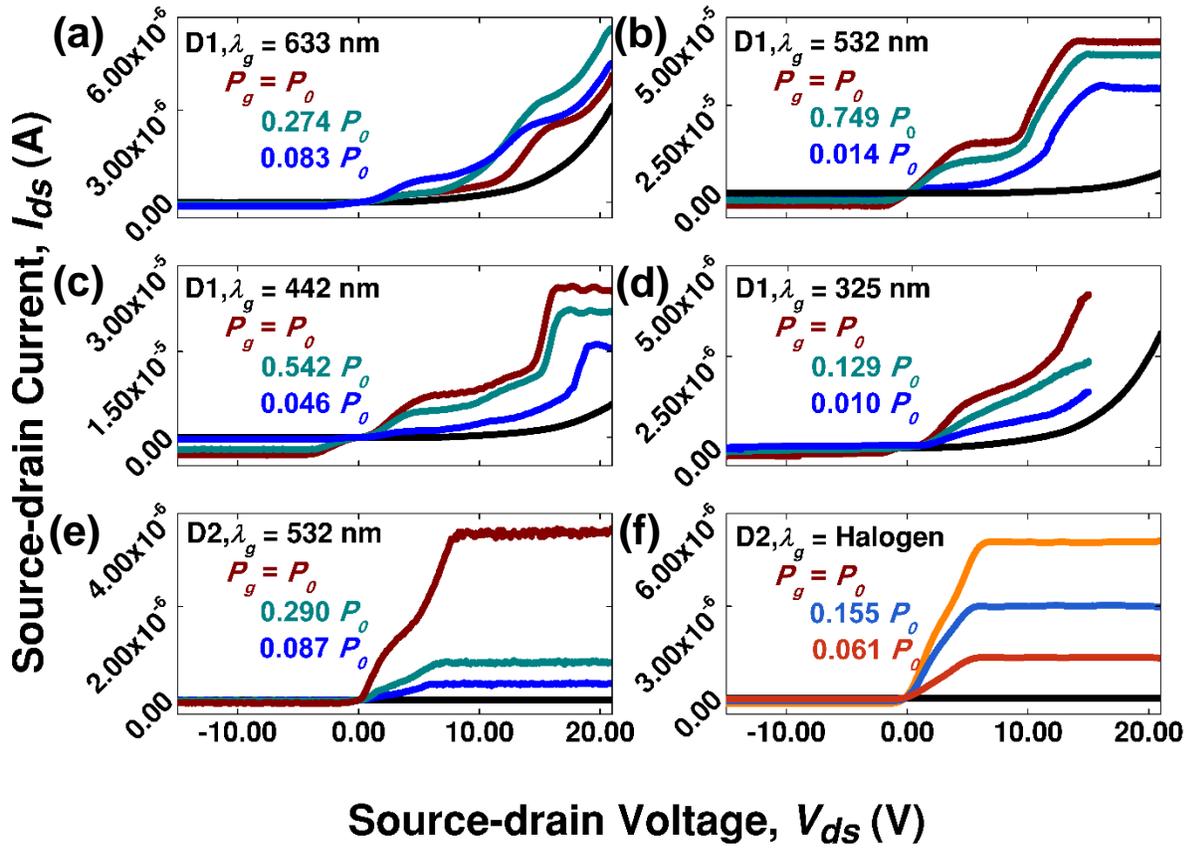

**Figure 3** LET output characteristics: Source-drain current, $I_{ds}$, as a function of the applied source-drain voltage, $V_{ds}$ with varying gate power, $P_g$, and wavelength, $\lambda_g$, for two devices (D1 and D2). (A-D) are for D1 under 633, 532, 442, and 325 nm illumination with $P_0$ values of 1.40, 2.07, 2.38, and 2.25 µW respectively, while (e-f) are for D2 under 532 nm and halogen excitation with respective $P_0$ values of 1.38 and 69.1 µW. The dark current is represented as black lines.

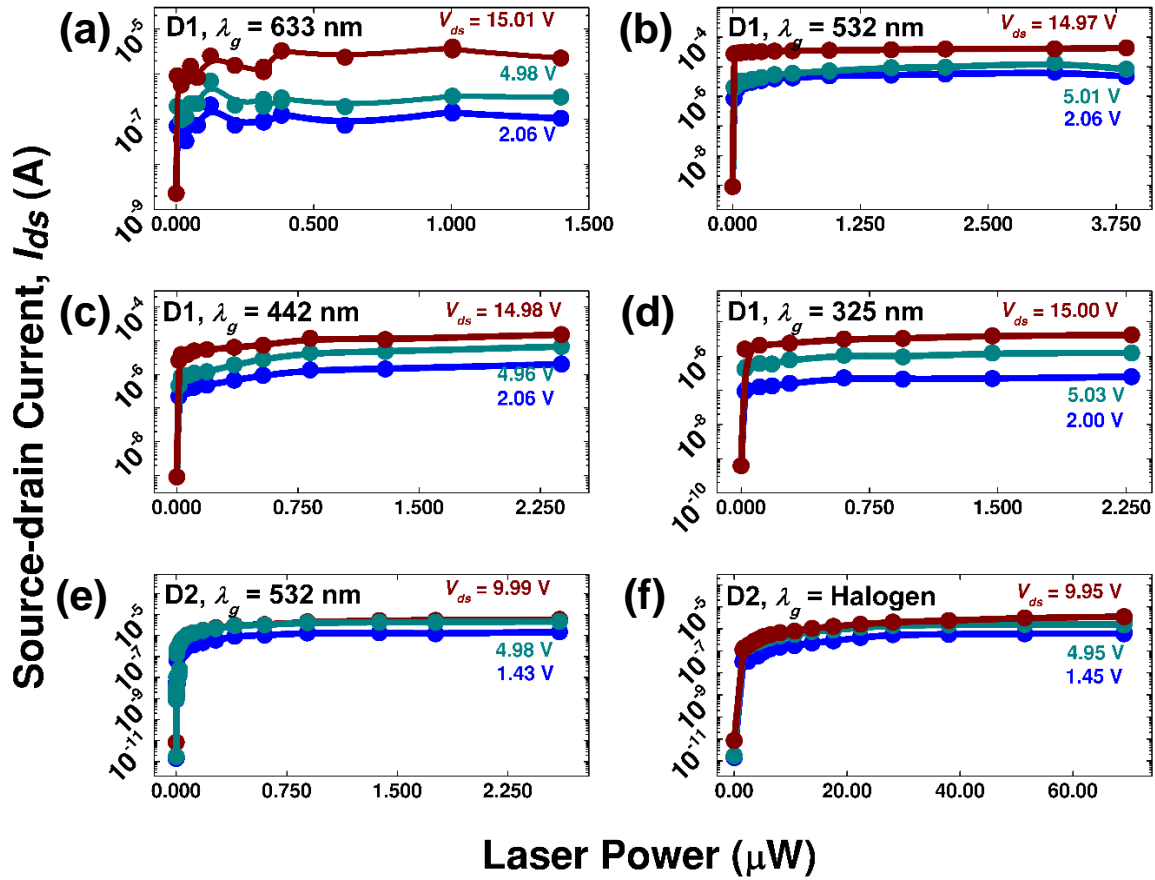

**Figure 4** LET transfer characteristics: Source-drain current, $I_{ds}$, as a function of laser power under different source-drain voltages, $V_{ds}$. **(a-d)** are for device D1, while **(e-f)** are for device D2 using the same conditions as in Fig. 3.

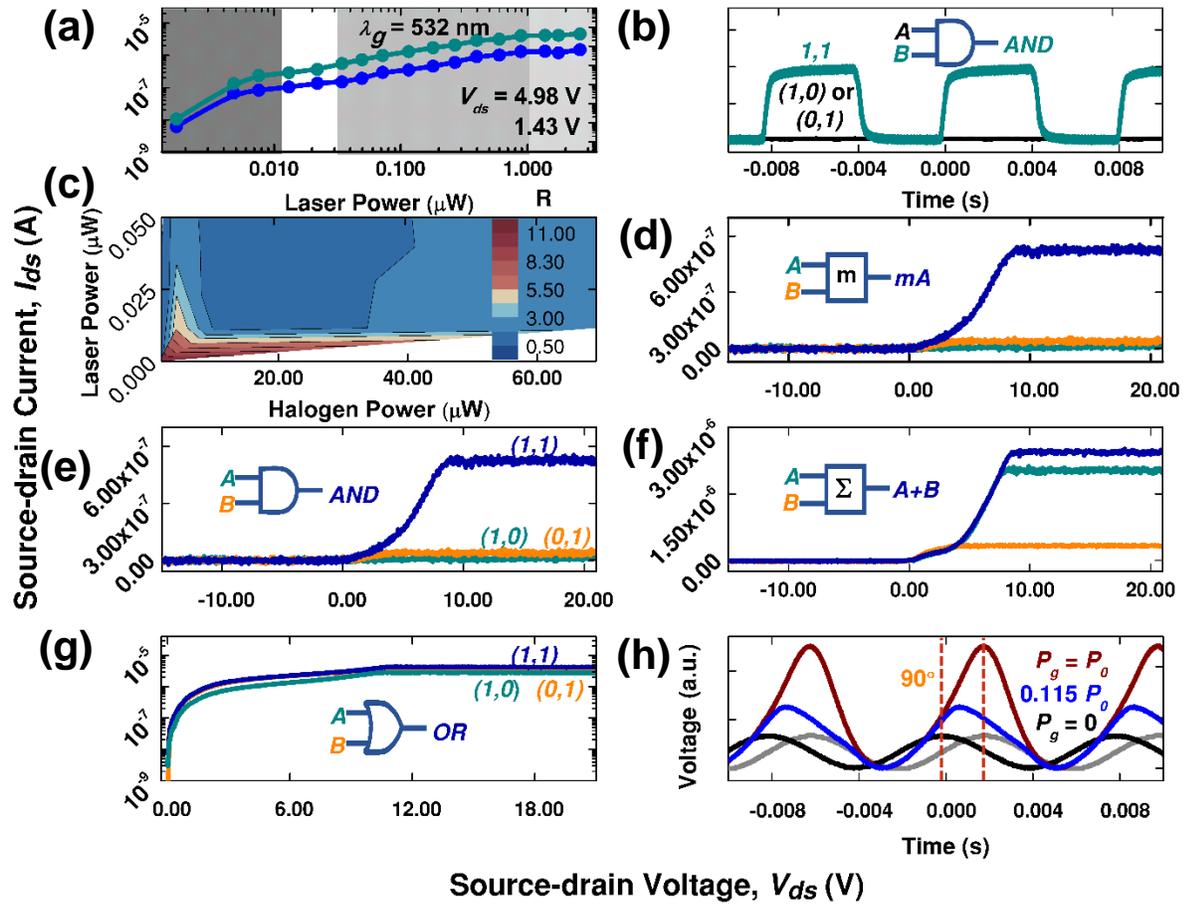

**Figure 5** Selected LET functionality demonstrations using D2. The axes are source-drain current, $I_{ds}$, vs. source-drain voltage, $V_{ds}$, except where noted otherwise. (**a**) Log-log plot of $I_{ds}$ vs. $P_g$ curves under 532 nm excitation with $V_{ds}$ values of 1.43 and 4.98 V, where the three shaded areas are visual guides for distinguishing the super-linear (dark gray), linear (medium gray), and saturation (light gray) regions used to demonstrate LET behaviors and applications in (b-h). (**b**) Optically modulated *AND* logic gate where $A = V_{ds}$ (5.00 V) and $B =$ modulated $P_g(532nm)$ (up to 2.60 μW in amplitude). Dark line: $V_{ds} = 5$ V and $P_g = 0$; green line: $V_{ds} = 5$ V and $P_g$ is modulated. (**c**) Various operation regions, according to ratio $R$ (see text for definition), achievable with two-beam illumination under a fixed $V_{ds}$ of 5.0 V. (**d**) A typical $R \gg 1$ operation point, with $P_{g1}(532nm) = 2$ nW and $P_{g2}(halogen) = 1.57$ μW, used as a demonstration of optical

amplification; and **(e)** contains the same data as (d) but used as a demonstration of an optical AND gate instead. **(f)** A typical $R \approx 1$ operation point, with $P_{g1}(532nm) = 0.63$ µW and $P_{g2}(halogen) = 0.7$ µW, used as a demonstration of summation operation. **(g)** A typical $R \approx ½$ operation point, with $P_{g1}(532nm) = 0.63$ µW and $P_{g2}(halogen) = 69.1$ µW, used as an optical *OR* logic gate. **(h)** LET operation under electrical modulation of $V_{ds}(t)$, while varying $P_g$ ($P_0 = 2.60$ µW). The outputs $I_{ds}(t)$ were measured through a sampling resistor. The input is shown in gray (normalized to the black $P_g = 0$ output curve).

**Supporting Information for**

**Light-effect transistor (LET) with independent gating for optical logic gates and optical amplification**

Jason K. Marmon[1-3], Satish C. Rai[4], Kai Wang[4], Weilie Zhou[4], and Yong Zhang[1-3]

[1]Nanoscale Science Program, University of North Carolina at Charlotte, Charlotte, NC 28223, USA.

[2]Department of Electrical and Computer Engineering, University of North Carolina at Charlotte, Charlotte, NC 28223, USA.

[3]Center for Optoelectronics and Optical Communications, University of North Carolina at Charlotte, Charlotte, NC 28223, USA.

[4]Advanced Materials Research Institute, University of New Orleans, New Orleans, LA 70148, USA.

Correspondence and requests for materials should be addressed to Y.Z. (e-mail: yong.zhang@uncc.edu) and W.Z. (e-mail: wzhou@uno.edu).

## Supplementary Materials Summary:

**S1. Estimated Actual Power Absorption**

Estimation of the laser spot size, and its area relative to the nanowire investigated to determine the actual power absorbed under different wavelengths.

**S2. Nanowire Synthesis & Device Fabrication**

**S3. Optical & Electrical Measurements**

**S4. Collection Efficiency: Uniform and Focused Illumination**

Quantum efficiency plots as a function of electron lifetime and nanowire length after Mott and Gurney[22] to demonstrate potential output characteristic changes from altering the optical gate or illumination position along the nanowire.

**S5. LET Transfer Characteristics with 532 nm and Halogen Illumination**

Two-beam point plots with data used to create contour plots in the main text.

**S6. Non-linear Dual Beam LET Transfer Characteristics with 633 nm and Halogen Illumination**

A two-beam illumination contour plot under another wavelength (633 nm instead of 532 nm) that demonstrates non-linear behavior.

**S7. Proposed Truth Tables and Symbols for AND-AND and AND-OR Logic Gates**

## S1. Estimated Actual Power Absorption

The laser spot size is estimated by the optical diffraction limit formula $1.22\lambda/N.A.$, where *N.A.* is the numerical aperture of the microscope lens. The fraction of the laser power actually absorbed is estimated by taking the ratio of the nanowire diameter to the laser spot size. The estimated ratios for the 632.8, 532, 441.6, and 325 nm lasers are 5.18, 6.16, 7.43, and 10.1% for a nanowire with an 80 nm diameter (device D1). For halogen illumination, the fraction of actual absorbed light is estimated using the ratio of the nanowire's cross section to the total illumination area. For the 50x LWD (10x MPLAN) objective lens, the illumination area is ~279 (~1450) µm$^2$. The ratio for the 80 nm wide/10 µm long nanowire (D1) is ~3.2 10$^{-6}$, and the power estimation for light actually absorbed is ~0.22 µW (which is comparable to that for the focused laser beam). All the illumination powers mentioned in the manuscript were applied powers, unless an actually absorbed power was explicitly stated.

## S2. Nanowire Synthesis & Device Fabrication

CdSe nanowires were grown in a vertical array through gold-catalyzed chemical vapor deposition, as described elsewhere,[27] and were then dispersed in alcohol and drop caste onto a Si substrate coated with a 300-nm thick $SiO_2$ layer or $Si/SiO_2$ chip. After CdSe nanowires were dispersed onto a chip, a thin poly-methyl methacrylate (PMMA) layer was spin coated onto the chip, followed by electron-beam lithography to open channels at a nanowire's ends. Exposed PMMA was removed by developing the chip. Afterwards, the chip was transferred to a thermal evaporator (Cressington-308R) for indium metallization (30 nm), followed by lift-off in acetone to obtain a finished device. The other indium wire end was bonded to a large gold pad used for placement of a gold-coated electrical probe. The samples were air stabilized for at least a week prior to testing.

## S3. Optical & Electrical Measurements

$I_{ds}$ vs. $V_{ds}$ measurements were collected with a Keithley© 2401 low voltage sourcemeter® that was remotely operated with LabTracer v2.9 software via a GPIB connection. For currents below ~1 nA, a Stanford Research System SR570 current pre-amplifier was used in conjunction with the Keithley©. Illumination sources consisted of halogen light, 532.016, 441.6, and 325 nm lasers ported through a Horiba LabRAM HR800 confocal Raman system with an internal 632.8 nm laser. Due to limited probe spacing for electrical measurements, all illumination sources were focused through a 50x long working distance (LWD) objective lens ($N.A. = 0.50$), except 325 nm, which went through a 10x MPLAN objective lens ($N.A. = 0.25$). Laser powers were limited to absolute powers of ~3 µW, as measured on the sample side of the microscope lens, to avoid potential laser-induced material modifications. Laser powers were altered through a combination of a standard neutral density filter in the Raman system and an adjustable neutral density filter in the laser path. Laser powers were measured with a Thor Labs PM100D power meter, and six and ten averaged measurements were used for D1 and D2, respectively, to calculate average powers and their standard deviations. The total power of the halogen light was estimated to be 69.1 µW with the power.

## S3. Collection Efficiency: Uniform and Focused Illumination

In the photoconductive region that Mott & Gurney[22] refer to as "primary photoconductivity," the electron collection efficiency, $\Psi$, is equivalent to the quantum efficiency, $\eta_{QE}$, assuming 100% absorption efficiency, and can be described with either Eqs. S1 or S2 for focused or uniform illumination respectively. Focused (uniform) illumination is when a single point (the entire length) of the nanowire is illuminated.

$$\Psi_{focused} = \frac{w}{L}\left(1 - e^{-x_0/w}\right) \tag{S1}$$

$$\Psi_{uniform} = \frac{w}{L}\left[1 - \frac{w}{L}\left(1 - e^{-L/w}\right)\right] \tag{S2}$$

In Eq. S1, $x_0$ is distance from the illumination site to the anode electrode, $L$ is the conducting channel's length, and $w$ is the electron mean free path and is related to the applied electrical field, $F$, or voltage, $V$, through $w = vFT = vVT/L$. The variables $v$ and $T$ are the carrier mobility and carrier lifetime, respectively. Eqs. S1-S2 may be compared to provide additional insight into the two typical operation modes.

Figs. S1A-B plots $\eta_{QE}$ or focused (dashed lines) and uniform (solid lines) illumination conditions as a function of applied voltage with (A) varying $T$ in a realistic carrier lifetime range, e.g. 0.1-100 ns,[43] when $L = 5.5$ µm (the same as D2), and with (B) nanowire lengths varying from L = 0.1 to 10 µm, assuming $T = 1$ ns. For focused illumination, $x_0/L = 0.5$ is assumed. A typical bulk CdSe mobility value[43] of 660 cm$^2$ V$^{-1}$ s$^{-1}$ is used. Fig. S1 indicates that two illumination modes yield comparable efficiency, and as a result, either focused or broad-area illumination could be reasonably implemented into a device, depending on the specific application need. An upper efficiency limit of 50% exists as only electrons are considered. In Fig. S1A, increasing the electron's mean free path by increasing carrier lifetime resulted in reaching a higher efficiency more rapidly, which suggests the ability to control current production at low applied bias. Fig. S1B indicates that nanowire length may also control the low bias current, where shorter nanowire devices produce larger currents. D1 and D2 were ~10. and ~5.5 µm in length; therefore, the shorter D2 is expected to exhibit a higher low bias current as demonstrated in Fig. 3. Mott & Gurney's classical equation is able to qualitatively reproduce trends observed for both devices (D1 and D2) in the low bias region, but this model does not describe the device characteristics in the higher bias region.

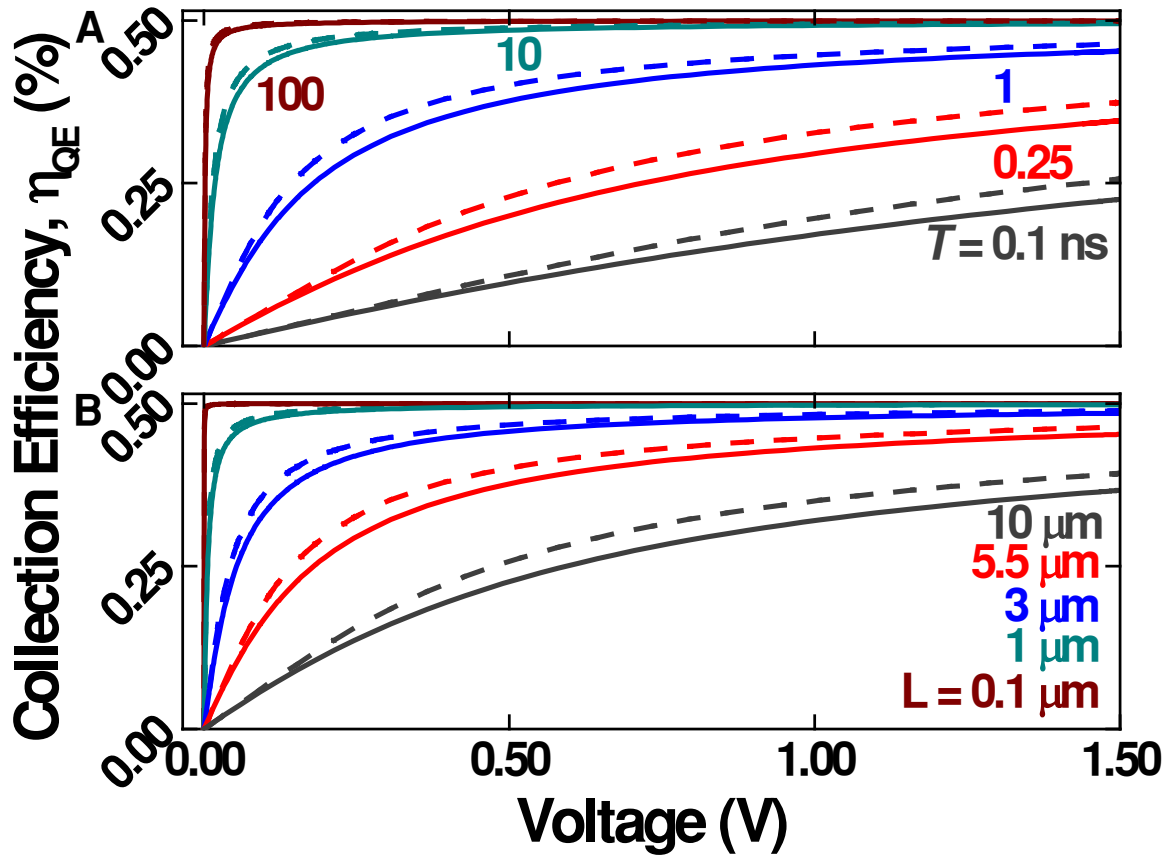

**Figure S1.** Electron collection efficiency, $\psi$, or quantum efficiency, $\eta_{QE}$, after Mott & Gurney,[22] as a function of applied bias for (A) different electron lifetimes, $T$, when $L = 5.5$ µm, and for (B) different nanowire lengths, $L$, when $T = 1$ ns, was plotted for focused illumination with $x_0 / L =$

0.5 (dashed lines) and uniform illumination (solid lines) conditions as described by Eqs. S1 and S2 respectively (see Supplementary Discussion). Lengths of 5.5 and 10 µm correspond to SNW lengths in D2 and D1 respectively. Note that this theory does not account for dark current differences between the two devices.

**S3. LET Transfer Characteristics with 532 nm and Halogen Illumination**

LET transfer characteristics under two-beam (532 nm and halogen) illumination. Data for Fig. 5C (text) is provided (Fig. S2A). Additional data (Fig. S2B) are provided to clearly show extension of three operating regimes observed for single beam illumination to dual beam illumination.

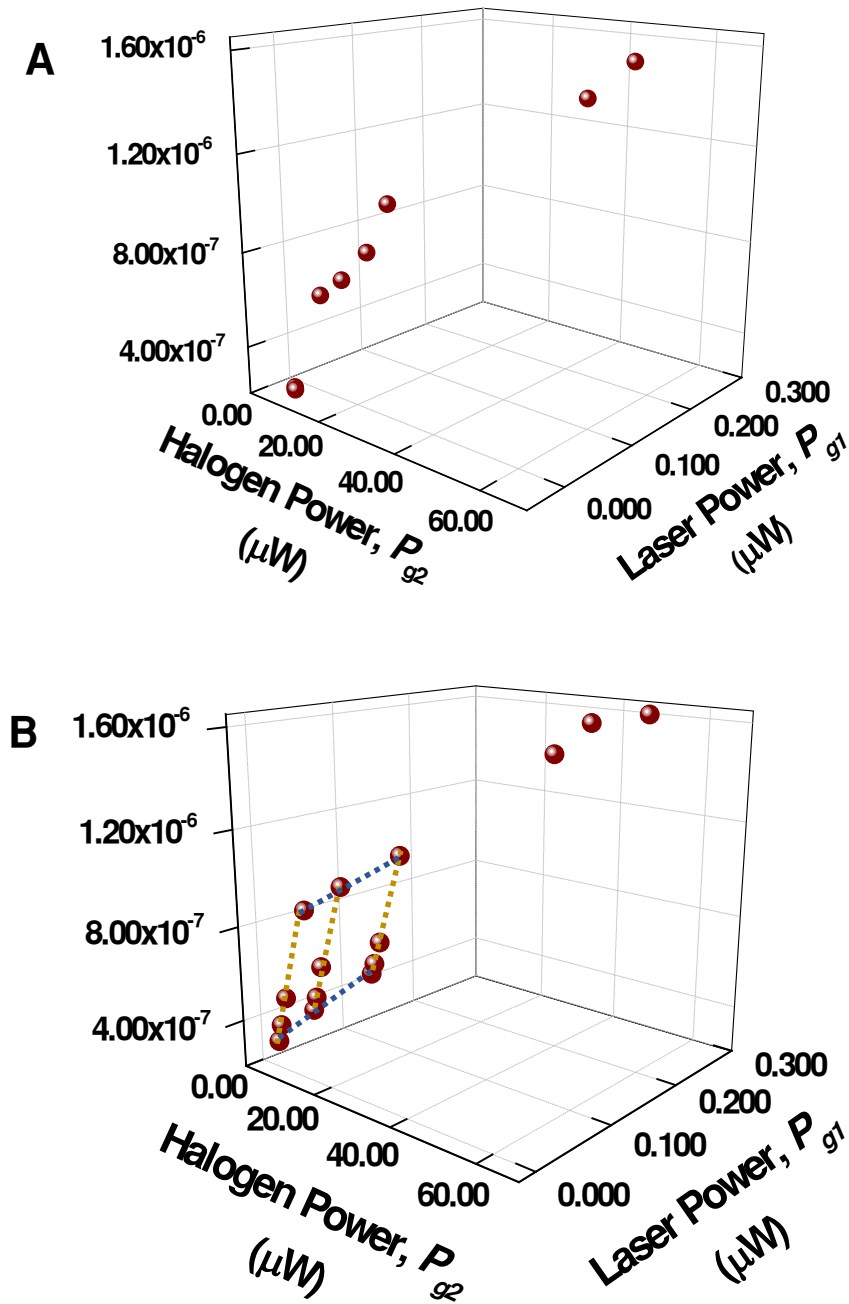

**Figure S2.** Dual gate LET transfer characteristics for $P_{g1}(532nm)$ and $P_{g2}(halogen)$ at $V_{ds} = 4.98$ V. The vertical axis is $I_{ds}$, and the horizontal axes are $P_{g1}(532nm)$ and $P_{g2}(halogen)$ respectively. A 3D plot of (A) the original $I_{ds}$ vs. $(P_{g1}, P_{g2})$ used to generate $R$ in Fig. 5C's contour plot, and (B) additional $I_{ds}$ vs. $(P_{g1}, P_{g2})$ data clearly demonstrating superlinear, linear,

and saturated regimes. The data in A consists of only three (four) measurements under combinations of 532 nm (halogen) illumination, which makes the three regimes more difficult to distinguish, but was chosen for Fig. 5 due to high $R$ values. B is the two-beam equivalent to the three operating regimes for single beam illumination shown in Fig. 4. The dotted lines guide the eye in B: the yellow line represents superlinear operation where $P_{g2}$ (halogen) is varied with fixed $P_{g1}$ (532 nm), while the light blue line illustrates linear operation where $P_{g1}$ is altered under constant $P_{g2}$. Notably, halogen illumination created a far greater current change than by altering the laser power; this is a direct result of the laser powers explored and suggests a wide range of customizable behavior solely through changing the illumination power ranges. The higher $(P_{g1}, P_{g2})$ points, e.g. $P_{g2} > 0.200\ \mu\text{W}$, are in the saturation regime.

**S4. Non-linear Dual Beam LET Transfer Characteristics with 633 nm and Halogen Illumination**

Contour plot for dual-beam LET (D2) illumination demonstrating non-linear behavior using 633 nm and halogen beams. $R$ is defined in the text.

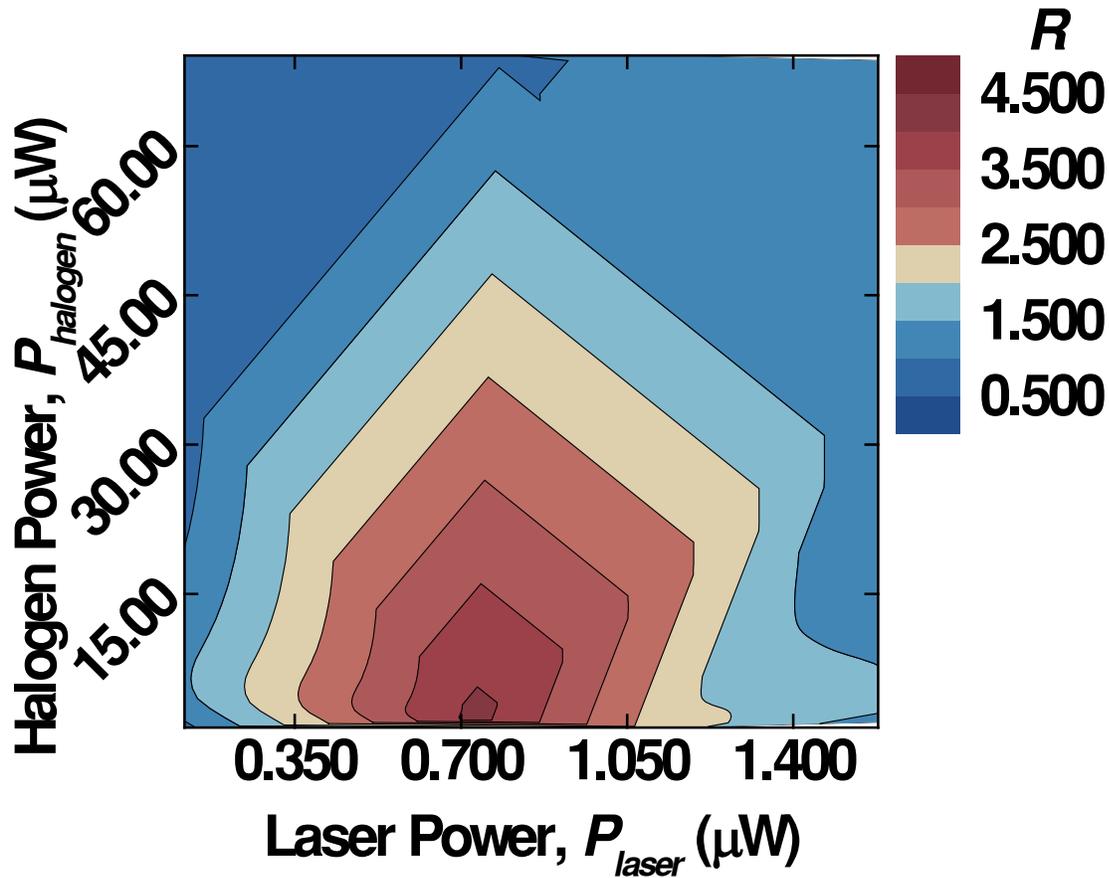

**Figure S3.** Dual-gate LET transfer characteristics for $P_{g1}(633nm)$ and $P_{g2}(halogen)$ at $V_{ds} = 4.98$ V. Two beam conditions produced non-monotonical behavior for both D1 (Fig. 3A in main text), and for D2 (this figure's data with $P_{g2}(halogen)=0$). For convenience, we represent this behavior as a contour plot. The non-linear behavior produced by 633 nm illumination allows for additional, interesting functionality beyond that offered by shorter wavelengths, e.g. 532 nm. Instead of switching between different laser powers, one only needs to modulate $V_{ds}$. For example, modulating $V_{ds}$ at a constant power of 1.06 µW allows dynamic switching between all three regimes or functionalities. This, in principle, simplifies the operational principle.

## S4. Proposed Truth Tables and Symbols for AND-AND and AND-OR Logic Gates

Truth tables (A and B) and their proposed symbols (C-D) for electrical-optical hybrid three-input *AND* and *AND-OR* logic gates (see text).

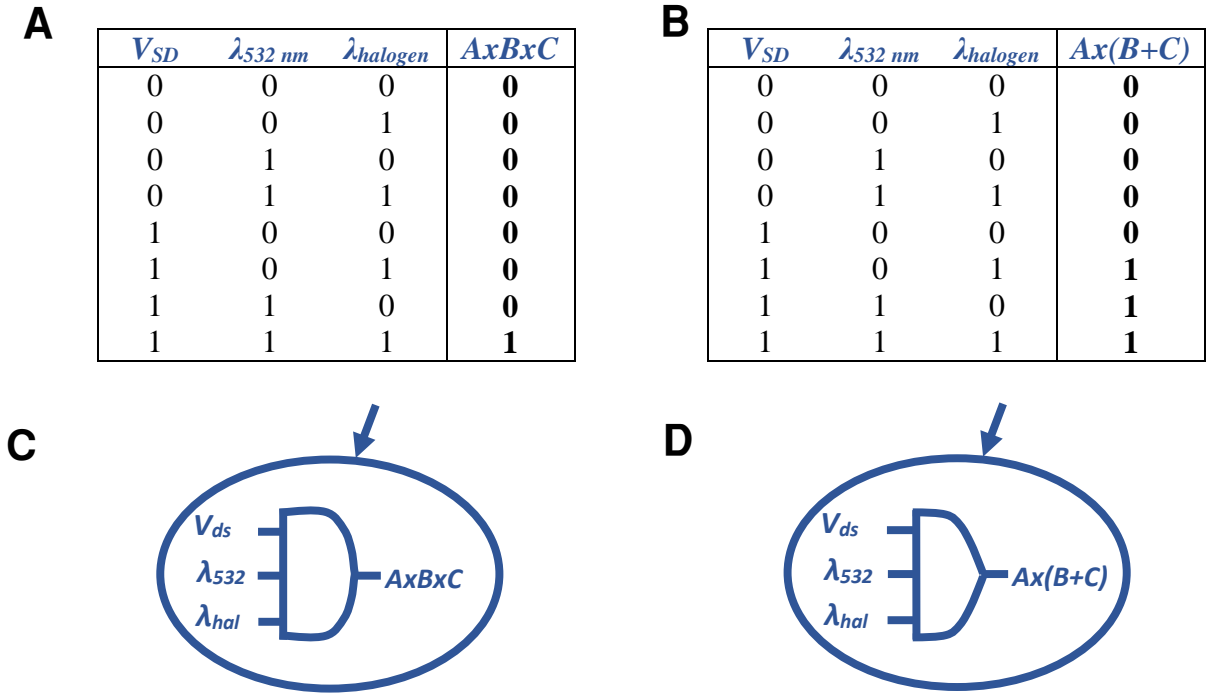

| $V_{SD}$ | $\lambda_{532\,nm}$ | $\lambda_{halogen}$ | $AxBxC$ |
|---|---|---|---|
| 0 | 0 | 0 | 0 |
| 0 | 0 | 1 | 0 |
| 0 | 1 | 0 | 0 |
| 0 | 1 | 1 | 0 |
| 1 | 0 | 0 | 0 |
| 1 | 0 | 1 | 0 |
| 1 | 1 | 0 | 0 |
| 1 | 1 | 1 | 1 |

| $V_{SD}$ | $\lambda_{532\,nm}$ | $\lambda_{halogen}$ | $Ax(B+C)$ |
|---|---|---|---|
| 0 | 0 | 0 | 0 |
| 0 | 0 | 1 | 0 |
| 0 | 1 | 0 | 0 |
| 0 | 1 | 1 | 0 |
| 1 | 0 | 0 | 0 |
| 1 | 0 | 1 | 1 |
| 1 | 1 | 0 | 1 |
| 1 | 1 | 1 | 1 |

**Figure S4.** Truth tables (A and B) and their proposed symbols (C-D) for hybrid electrical-optical hybrid three-input *AND* and *AND-OR* logic gates, respectively, which are achieved using three inputs consisting of $A = V_{ds}$, $B = P_{g1}(532nm)$, and $C = P_{g2}(halogen)$. (A) *three-input AND* and (B) *AND-OR* logic gates have respective outputs of $AxBxC$ and $Ax(B+C)$. Changing between two logic functions simply requires altering both optical gate powers from the super-linear region to the saturation region in Fig. 5A. This example may be extended to other optical gate wavelengths. Logic gate symbols are composed of common electronics symbols. For example, the three-input *AND* gate uses the standard symbol, while the *AND-OR* gate is a mixture of *AND* and *OR* gates, respectively. The circle with arrow is commonly used to denote photo-active processes (similar to a photo-diode).

**References** (in addition to those in the main text)